\documentclass[times,twocolumn,final,authoryear]{elsarticle}

\usepackage{jasr}
\usepackage{framed,multirow}

\usepackage{amssymb}
\usepackage{latexsym}

\usepackage[bookmarks = true,colorlinks = true,linkcolor = blue,urlcolor = blue,citecolor = blue,bookmarks = true,linktocpage = true,hyperindex = true,breaklinks]{hyperref}

\usepackage{url}
\usepackage{xcolor}
\definecolor{newcolor}{rgb}{.8,.349,.1}

\usepackage{sidecap} 

\chardef\us=`\_

\journal{Advances in Space Research}

\begin{document}

\verso{Pooja Devi \textit{etal}}

\begin{frontmatter}

\title{Prominence Oscillations activated by an EUV wave \tnoteref{tnote1}}%

\author[1]{Pooja Devi\corref{cor1}}
\cortext[cor1]{Corresponding author: }
\ead{setiapooja.ps@gmail.com}
\author[1]{Ramesh Chandra}
\author[1]{Reetika Joshi}
\author[2]{P. F. Chen}
\author[3,4]{Brigitte Schmieder}
\author[5]{Wahab Uddin}
\author[6]{Yong-Jae Moon}
\address[1]{Department of Physics, DSB Campus, Kumaun University, Nainital-263001, India}
\address[2]{School of Astronomy \& Space Science, Nanjing University, Nanjing 210023, China}
\address[3]{Observatoire de Paris, LESIA, UMR8109 (CNRS), F-92195 Meudon Principal Cedex, France}
\address[4]{Centre for mathematical Plasma Astrophysics, Dept. of Mathematics, KU Leuven, 3001 Leuven, Belgium}
\address[5]{Aryabhatta Research Institute of Observational Sciences (ARIES), Nainital 263 001, India}
\address[6]{School of Space Research, Kyung Hee University, Yongin, Gyeonggi-Do, 446-701, Korea}

\begin{abstract} 
Prominence oscillations are 
one of interesting phenomena in the solar atmosphere, which can be utilized to infer the embedded magnetic field magnitude.
We present here the transverse oscillations of two different prominences located at the East solar limb on 2011 February 11 using the multi-wavebands data of the Atmospheric Imaging Assembly (AIA) on-board the Solar Dynamics Observatory (SDO) satellite. A prominence eruption was observed towards the east direction with an average speed of $\approx$ 275 km s$^{-1}$. The eruption is fitted with the combination of a linear and an exponential functions of time. An extreme ultraviolet (EUV) wave event was associated with the prominence eruption. This EUV wave triggered the oscillations of both prominences on the East limb. We computed the period of each prominence using the wavelet analysis method. The oscillation period varies from 14 to 22 min.  The magnetic field of the prominences was derived, which ranges from 14 to 20 G.
\end{abstract}

\end{frontmatter}

\section{Introduction}
\label{sec_introduction}


Solar prominences, also known as filaments on the solar disk, are high density cool materials hanging in the solar corona \citep[for reviews see:][]{Labrosse2010, Chen2020}. They are observed in various chromospheric and coronal filters in quiet regions as well in active regions. They can erupt partially or fully, whenever the balance between magnetic tension and magnetic pressure gradient is disturbed. The eruption of  prominences, either  full or partial, is often accompanied by solar flares from medium to large classes \citep[for example,][]{Aulanier2000, Liu2009, Schmieder2015, Joshi2017, Roudier2018}. When an eruption fails, sometimes the filament may return to a quasi-static equilibrium later on \citep[for example,][]{Joshi2013}.

An interesting phenomenon associated with prominence eruption is their association with extreme ultraviolet (EUV) waves \citep{Zheng2012, Chandra2016, Zheng2020}. The EUV waves are defined as the propagating bright fronts, first discovered by the Extreme ultraviolet Imaging Telescope (EIT) on board the Solar and Heliospheric Observatory (SOHO) satellite. Initially, they were explained as fast-mode magnetohydrodynamic (MHD)  waves \citep[][and references cited therein]{Thompson1998, Wang2000, Wu2001, Ofman2002, Vrsnak2002, Srivastava2016}. However, the finding of stationary fronts and the extremely slow speeds of some EUV waves could not be explained by the fast-mode MHD wave model \citep{Delannee1999, Delannee2000}. 
Therefore, to reconcile these discrepancies, \cite{Chen02} and \cite{Chen05} proposed a hybrid model. According to their model, EUV waves have two components, a non-wave (known as EIT wave) and a fast-mode MHD wave.
Sometimes, the fast mode EUV wave can be converted into a slow-mode wave when it interacts with a magnetic separatrix \citep{Chen2016}.
Later SDO observations confirmed this model in many events \citep{Zhukov2004, Chen2011, Zhao2011, Chandra2021a}. 

EUV waves phenomena can trigger oscillations of nearby or distant prominences/filaments, which have been observed for a long time \citep{Thompson1991, Zhang1991, Vrsnak1993, Shen2014L}. Prominence oscillations provide a useful tool to diagnose the coronal magnetic field \citep[see the review by][]{Arregui2018}. These oscillations are categorized into two types, namely, longitudinal and transverse.  
In longitudinal oscillations, the direction of  the oscillations is parallel to the prominence threads whereas in transverse oscillations, it is normal to the prominence threads \citep{Zhou2018}, not to the prominence spine as claimed by many papers.
Longitudinal oscillations are often excited by nearby impulsive events, for example, microflares and small jets \citep{Luna2012, Liakh2020, Zhang2020, Luna2021}, whereas transverse oscillations are often triggered by energetic disturbances produced by distant flares, EUV waves and CMEs \citep{Okamoto2004, Liu2012}, providing a useful tool to diagnose the coronal magnetic field. According to previous studies, longitudinal oscillations usually have a longer period in comparison to transverse oscillations \citep{Luna2014}. It is also reported that longitudinal and transverse oscillations can be observed simultaneously in separate filaments or in a single filament \citep{Shen2014L, Pant2016, ZhangLiDing2017, Zhang2017,Mazumder2020, Zhang2020}.

The restoring forces  acting for these two kinds of oscillations are quite different.
In the longitudinal oscillations, gravity is the main restoring force \citep{Luna2012, Zhang2012}. Using 1-dimensional hydrodynamic numerical simulations, \citet{Luna2012} and  \citet{Zhang2012} demonstrated that the field-aligned component of gravity can explain the obtained period, favoring the pendulum model. Later on, \citet{Luna2016} did 2-dimensional (2D) MHD simulations and concluded that it is similar to the simplified pendulum model. 2D non-adiabatic MHD simulations  were also performed  by \citet{Zhang2019} to explain the longitudinal oscillations. Theoretical models for transverse oscillations in filaments have been provided by \citet{Hyder1966} and \citet{Kleczek1969}. In these models, the magnetic tension is considered as the main restoring force. 

In this article, we present the transverse oscillations of the prominences situated at the east limb. These oscillations were triggered by an EUV wave associated with a filament eruption on 2011 February 11. The article is organized as follows: Section \ref{sec_obs} presents the observational data sets, event overview, the prominence oscillations and coronal seismology. In Section \ref{sec_summary}, the discussion and summary of the study are given.

\begin{figure}
\centering
\includegraphics[scale=0.48]{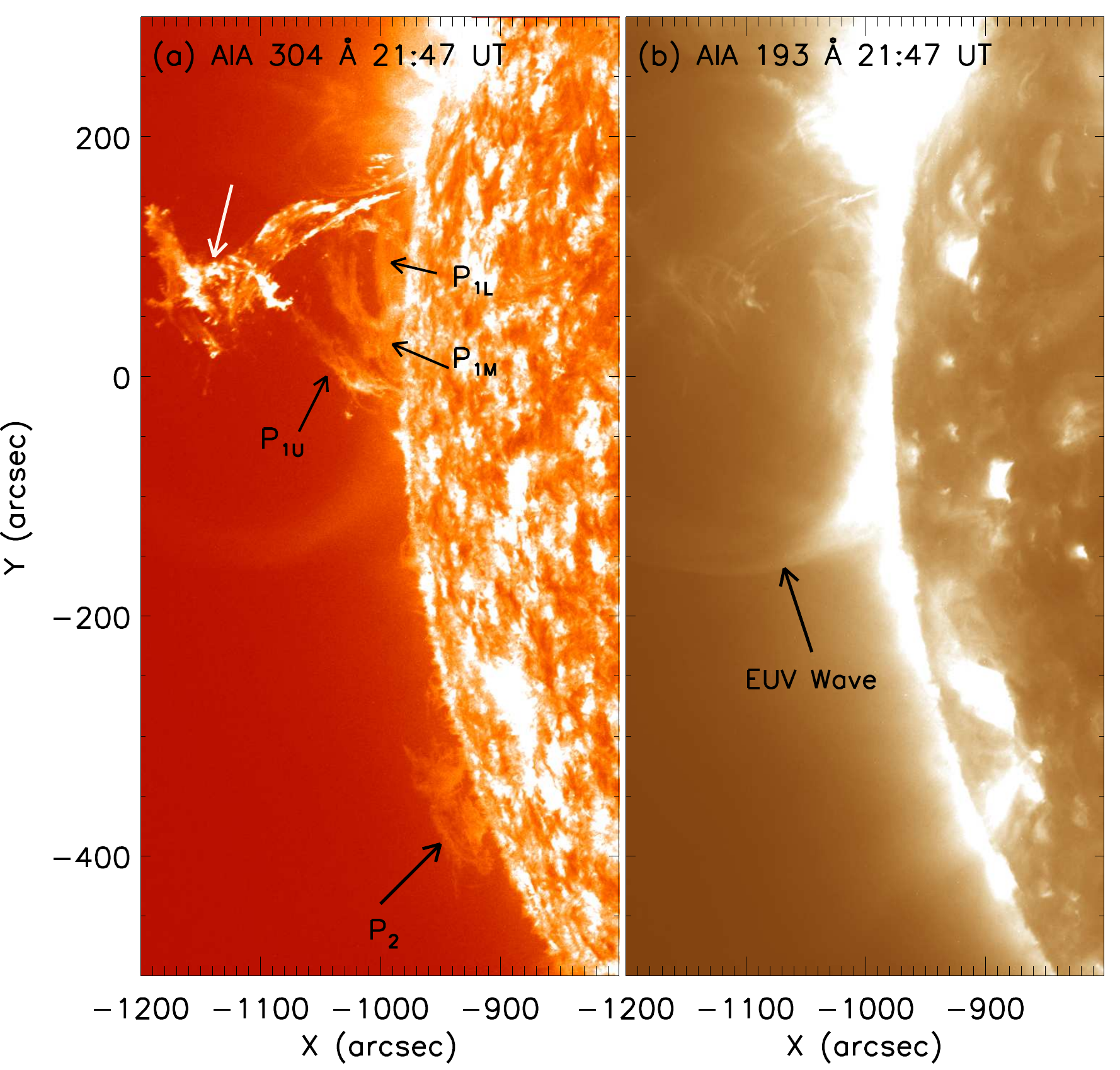}
\caption{Images of the prominences and EUV waves at AIA 304 and 193 \AA~ in panels (a) and (b), respectively. Three prominence layers P$_{1L}$, P$_{1M}$, P$_{1U}$, and another prominence P$_{2}$ are shown by the black arrows in panel (a). A white arrow in panel (a) presents the erupting prominence. The EUV wave is shown in panel (b). }
\label{fig_overview}
\end{figure}

\begin{figure*}[t]
\centering
\includegraphics[scale=0.63]{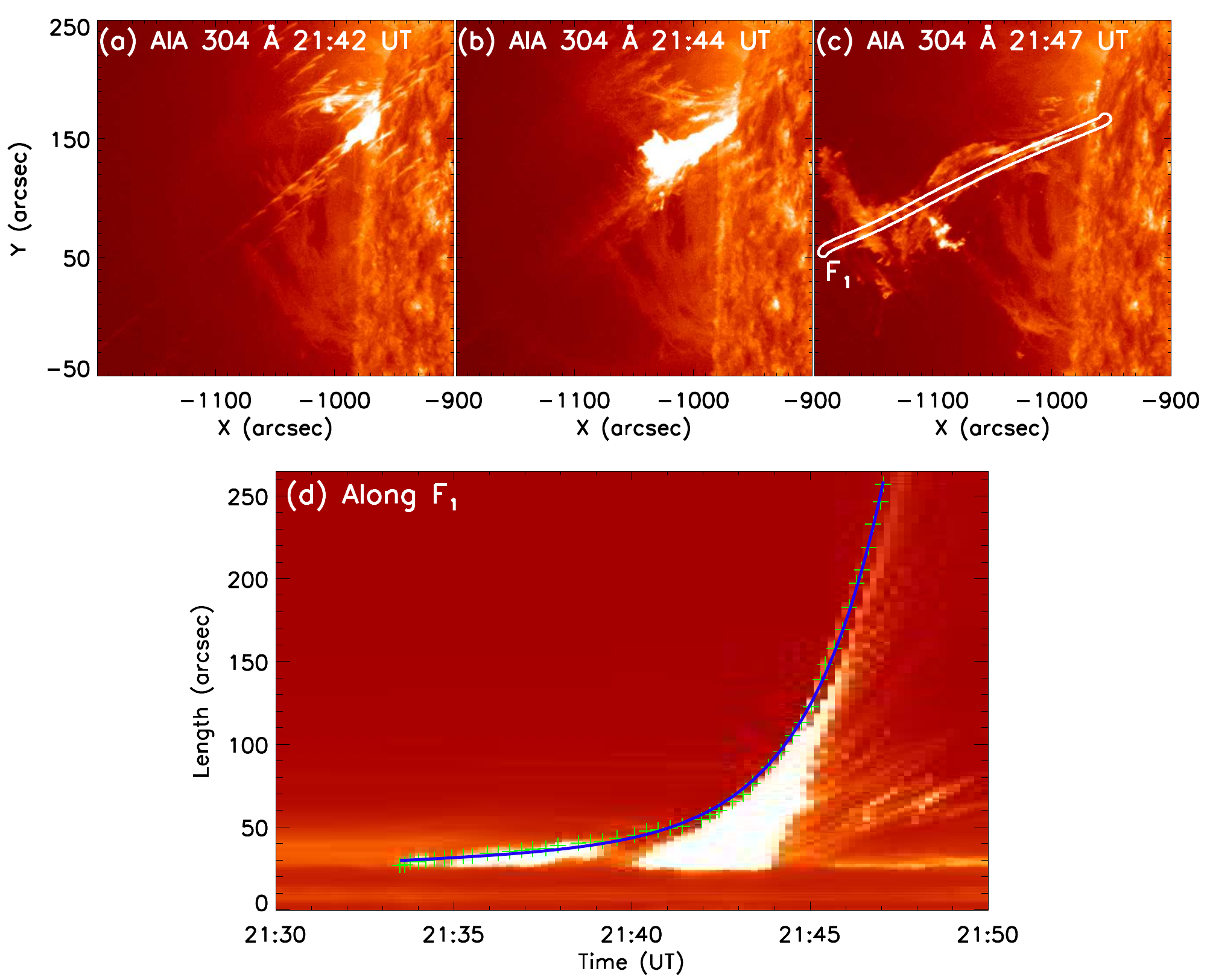}
\caption{Images showing the sequence of the prominence eruption in panels (a)--(c). The slice F$_{1}$ in panel (c) is chosen along the direction of prominence eruption, and panel (d) displays the time-distance diagram of the AIA 304 \AA\ intensity along the slice F$_1$. We fit a function ``$ a e^{b(t-t_0)}+ct+h_0 $'' in the leading edge of the eruption. These data points selected for the fitting are shown by green ``plus'' symbols whereas the blue solid line is the fitting of the function in the selected data points.}
\label{fig_evolution}
\end{figure*}

\section{Observations}
\label{sec_obs}
\subsection{Data sets}
The oscillations were observed in  most of the filtergrams of the Atmospheric Imaging Assembly \citep[AIA,][]{Lemen2012} on board the Solar Dynamics Observatory \citep[SDO,][]{Pesnell2012}. Considering the visibility, in the current study, we used the data in AIA 304 \AA\  and 193 \AA. The spatial resolution and the cadence of the AIA data are 1.2$''$ and 12 sec, respectively. The data are processed using  different routines available in the solar software (SSW).

\subsection{Event Overview}
On 2011 February 11, at the east limb a prominence eruption was observed and the eruption was associated with a GOES B8.0-class flare. The prominence eruption was initiated at $\approx$ 21:40 UT in the east direction and later on it evolved to a slow coronal mass ejection (CME) at 22:12 UT in the Large Angle and Spectrometric Coronagraph (LASCO) C2 field-of-view with a linear average speed of $\approx$ 469 km s$^{-1}$.
In Figure 1, the erupting and the oscillating prominences are shown in the left panel with white
and black arrows, respectively, where P$_{1L}$, P$_{1M}$, and P$_{1U}$ stand for the lower, middle, and upper
layers of the first prominence, respectively. P$_2$ is the second oscillating prominence located in a 
different active region. The EUV wave is highlighted in the right panel.

The evolution of the prominence eruption in AIA 304 \AA\ is presented in the upper panel of Figure~\ref{fig_evolution}. For the kinematics of the eruption, we selected an artificial slice in the eruption direction. The slice is drawn in panel (c) of the same figure and named as F$_1$. The time-distance plot along this slice is displayed in panel (d) of the figure. The combination of a linear and an exponential function is
used to fit the height evolution of the erupting prominence. The function is in the form of  $a e^{b(t-t_0)}+ct+h_0 $, where $a,~b$, $c$, and $h_0$ are constant parameters to be found by the fitting.
This function is similar to the function used by \citet{Cheng2020} and \citet{Chandra2021b}.
We fix $t_0$ to our first data point which is at 21:33:24 UT. The data points for the fitting are selected manually at the outer edge of the brightening around the slice. These data points are shown by green ``plus'' symbols and the blue solid line is the fitting of the function in the selected data points. 
The parameters $a$, $b$, and $c$ are determined in the fitting to be 64.46 km, 0.46 min$^{-1}$ , and 10.36 km $s^{-1}$, respectively.
The value of the reduced $\chi$-square for the current fitting is 1.6. 
The average speed of the eruption  is $\approx$ 275 km s$^{-1}$.

Apart from the erupting prominence, at the east limb we find another two prominences. One is close to the erupting prominence and contains three layers one above another (like decker configuration) and they are labelled as P$_{1L}$, P$_{1M}$, and P$_{1U}$, respectively. The second prominence was in the southern hemisphere and is labelled as P$_2$. The locations of these prominences in AIA 304 \AA\ along with the erupting prominence are depicted in Figure~\ref{fig_overview} (left panel). 

The eruption of the main prominence drove an EUV wave, which is shown in AIA 193 \AA\ in the right panel of Figure~\ref{fig_overview}. The details of this EUV wave event were studied by \citet{Chandra2021a} using the multi-wavelength and multi-view point observations. As discussed in \citet{Chandra2021a}, such an EUV wave event has two components i.e. a slower non-wave which is interpreted to be due to magnetic field line stretching and a fast-mode MHD wave. The speed of the slower wave is 298$\pm$5 km s$^{-1}$, whereas the fast-mode wave speed is 445$\pm$6 km s$^{-1}$. 

\subsection{Prominence Oscillations and Coronal Seismology}

\begin{figure*}[ht!]
\centering
\includegraphics[scale=0.68]{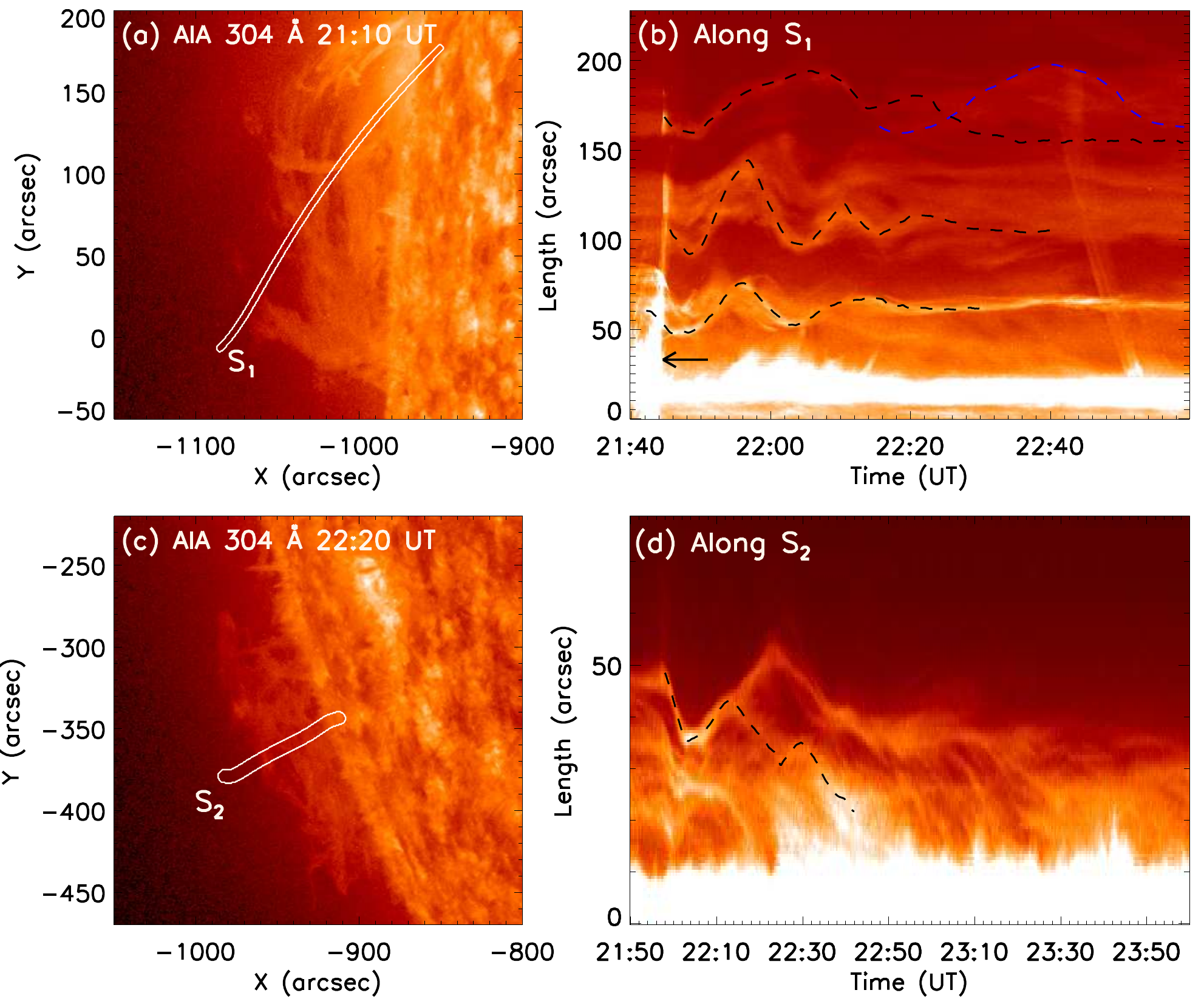}
\caption{Slices S$_1$ and S$_2$ along the directions of the oscillating prominences P$_1$ and P$_2$ are displayed in panels (a) and (c), respectively. Panels (b) and (d) show the time-distance diagrams along slices S$_1$ and S$_2$, respectively. A black arrow in panel (b) denotes the EUV wave crossing the prominence threads. The black dashed lines in panels (b) and (d) are the tracking of the prominence oscillations from the time-distance plots. The blue dashed line in panel (b) represents the peak, in which we did not find the complete oscillation. It could be another thread of the prominence different from the oscillating one traced by the black dashed line. Therefore, we are not considering this for our period analysis. An animation of the this figure is available online.} 
\label{fig-Osci}
\end{figure*}

When the fast component of the EUV wave, i.e. the fast-mode MHD wave, moved towards the east-south direction, it encounters the prominences situated on the east limb. As a result, the prominences  P$_{1L}$, P$_{1M}$, P$_{1U}$, and P$_2$ oscillated. The oscillations are clearly visible in the attached movie. 
In order to study the prominence oscillations in detail,
we select two slices shown in panels (a) and (c) of Figure~\ref{fig-Osci}. The time-distance plots along these slices are drawn in panels (b) and (d) of the figure, respectively. In panel (b), we can see the oscillations of the P$_{1L}$, P$_{1M}$, and P$_{1U}$ prominence layers. In the same way the oscillation of prominence P$_2$ is depicted in panel (d) of the same figure. We track manually the evolution of the center of each oscillating prominence, which is over-plotted as the black-dashed lines in the same figure.

\begin{figure*}
\centering
\includegraphics[scale=0.7]{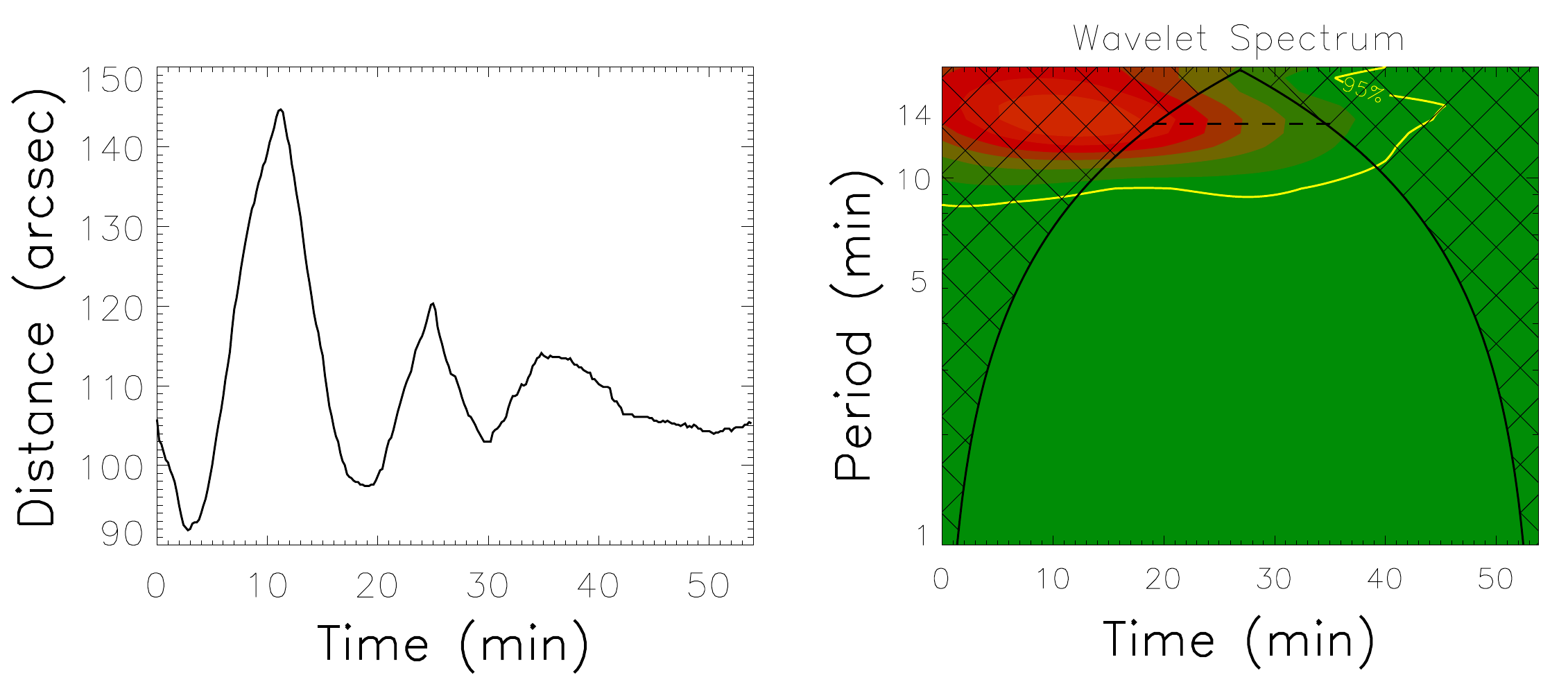}
\caption{Temporal evolution of the displacement of the prominence P$_{1M}$ (left panel). The right panel displays the wavelet spectrum, where the horizontal dashed line corresponds to the peak of global power spectrum at 14.0 min. The starting time is 21:50 UT.}

\label{fig-wavelet}
\end{figure*}

The above manually tracked points are used for the time-period analysis of these oscillations. To calculate the time period, we apply the wavelet analysis technique, with a significance test for the levels higher than or equal to 95\% as real (Figure \ref{fig-wavelet}). The details about the wavelet technique and significance test were well explained in \cite{Torrence1998}. 
The significant region is bounded by the cone-of-influence (COI) region, which acts as an important background for the edge effect for the given time range \citep{Luna2017,Joshi2020}. 
This method is used to find the period of all the four oscillations in P$_{1U}$, P$_{1M}$, P$_{1L}$, and P$_{2}$.
The derived periods of the oscillations vary from 14 to 22 min and are provided in Table \ref{table}.
An example of the wavelet analysis is given in Figure \ref{fig-wavelet}. In this figure, we display the time evolution of the displacement and its wavelet spectrum for the oscillations of prominence P$_{1M}$, where the black dashed line marks the peak period at 14.0 min. As for the amplitude of the oscillations in the two selected slices S$_1$ and S$_2$, we found that the amplitude is significantly weaker along slice S$_2$. The reason is straightforward, i.e., the slice S$_2$ is much further away from the eruption source region. As the EUV wave propagates out, it becomes weaker and weaker, hence the oscillation amplitude along slice S$_2$ is significantly smaller.

It is believed that the transverse oscillations are due to the dominant magnetic tension restoring force. Taking this fact into account, we consider the reported oscillations here like the kink-mode oscillation as suggested by \citet{Zhou2016} and \citet{Zhang2018}. Coronal seismology allows us to determine the physical properties of the prominences such as the magnetic field strength \citep{Nakariakov2005, Nakariakov2021}. We find the lengths of prominences by assuming that the prominences are along the magnetic field lines and the length of the coronal loops are approximately equal to prominence length. 
Then we could estimate the length of each prominence viewed as loops at  the limb, one over the other. We find that the length of the lower loop P$_{1L}$, middle loop P$_{1M}$, upper loop P$_{1U}$, and the prominence P$_2$ are $\approx$ 220, 260, 340, and 210 Mm, respectively. The magnetic field strength is calculated using the method proposed in  \citet{Ofman2018} and \citet{Shen2019}.
The kink speed of the oscillations is defined as $C_k~=~2L/P$, where $L$ and $P$ are the length of the magnetic field lines and period of the oscillation, respectively. The calculation gives the kink speed varying from 430 -- 620 km s$^{-1}$. The kink speed depends on the Alfv\'en speed as $C_k = V_A \sqrt{\frac{2}{1~+~n_o/n_i}}$, which is related to the magnetic field strength by the relation $V_A = \sqrt{\frac{B^2}{4\pi\rho}}$. Here, $n_o/n_i$ and $\rho$ are the ratio of number density outside/inside the prominence and the mass density of the prominence, respectively. Assuming the ratio $n_o$/$n_i$=0.01, we find the Alfv\'en speed in the range of 307 -- 441 km s$^{-1}$. It is noted that even $n_o$/$n_i$ increases by 10 times, i.e., $n_o$/$n_i$=0.1, the derived Alfv\'en speed would increase by only 4\%. Therefore, the influence of the density ratio on the result is negligible. From the calculated Alfv\'en speed and mean mass density, $\rho$ = 1.67$\times$10$^{-14}$ g cm$^{-3}$ from \citet{Labrosse2010}, the magnetic field strength in P$_{1L}$, P$_{1M}$, P$_{1U}$, and P$_2$ is found to be  14.4$\pm$1.3, 20.1$\pm$1.8, 16.3$\pm$1.6, and 14.4$\pm$1.2 G, respectively. All the calculated and derived parameters are listed in Table \ref{table}.

The prominence oscillation is due to stationary waves. In contrast, a traveling wave was also observed to propagate along the prominence layer $P_{1U}$ with an average velocity of $\approx$ 435 km s$^{-1}$ as indicated in the online animation of Fig. \ref{fig-Osci}, which provides another method to estimate the magnetic field of the prominence as demonstrated by \citet{Shen2019}.For this method, we assume that the average speed of the fast-mode EUV wave is of the order of Alfv\'en speed, i.e. $V_{avg} \sim V_A$ in the coronal conditions, where the sound speed is smaller than the Alfv\'en speed \citep[see][]{Shen2019}. Using this assumption, we find that the magnetic field strength of the prominences is about 20 G, which is close to the magnetic field strength computed by first method.
The difference in magnetic field measurement could be due to the uncertainties in the field-line lengths and the fast-mode wave speed measurements. 

Our estimated magnetic field strengths are comparable with the previously reported values based on coronal seismology or prominence seismology \citep{Vrsnak2007, DeMoortel2009, LiuW2012, Xue2014,Pant2015, Zhang2017, Luna2018, Mackay2020}. \cite{Luna2018} did a statistical study using the GONG H$\alpha$ observations of 196 solar filaments near the maximum of solar cycle 24 and they found that the minimum average magnetic field strength is about 16 G. Using the Hanle effect, the average field strength of prominences was measured to vary from 5 to 10 G \citep{Leroy1983, Leroy1984, Kuckein2009, Sasso2011, Kuckein2012, Xu2012, Raouafi2016}. It seems that the observed magnetic field strength using Hanle effect is slightly less than derived by coronal seismology.

\section{Discussion and Summary}
\label{sec_summary}
The present analysis explores the oscillations of prominences situated above the East limb on 2011 February 11. These oscillations are initiated by an EUV wave associated with the eruption of another prominence. Since the oscillations are normal to the prominence layers therefore these are  transverse oscillations. The main points of our study are as follows :

\begin{itemize}
\item{Both the prominence with three layers near the source eruption region and another prominence further south oscillate in response to the fast-mode EUV wave associated with the eruption.}
    \item{For the P$_{1L}$, P$_{1M}$, and P$_{1U}$, the periods are 17, 14, and 22 minutes, respectively. As expected, for the three layers of the first oscillating prominence, the period is the longest for the uppermost layer.}

    \item{We have computed the coronal magnetic field strength of the prominences with two different methods with comparable results. The derived value of the magnetic field strength ranges from 15 to 21 G.}
\end{itemize}

The oscillations in the solar atmosphere are commonly observed in different features like coronal loops, prominences, cavities, coronal streamers, etc. Mostly these oscillations are activated by external perturbations. EUV waves are one of the prominent causes for creating the oscillation in prominences \citep{Asai2012, Gosain2012}, cavities \citep{Zhang2018}, and coronal loops \citep{Ballai2007, Guo2015}. The present observations and analysis indicate that transverse oscillations in prominences  are triggered by a fast-mode EUV wave at the east limb. 
\citet{Asai2012} found the oscillations of a prominence and a filament triggered by an EUV wave as a fast-mode MHD wave with a velocity of about 570--800 km s$^{-1}$. \citet{Gosain2012} analysed a flare associated transverse oscillations in a quiescent prominence. \citet{Liu2012} and \citet{Zhang2018} reported on transverse oscillations in a cavity which are triggered by nearby EUV wave.

In the present case, we find the oscillations in the prominences P$_{1L}$, P$_{1M}$, P$_{1U}$, and P$_{2}$ by a single EUV wave (speed $\approx$ 445$\pm$6 km s$^{-1}$),  which hits these prominences during its propagation. The calculated periods are in range of 14 -- 22 min. 
This range is consistent with \citet{Asai2012}, \citet{Gosain2012}, \citet{Shen2014a}, and  \citet{Shen2014b}.
Further, the magnetic field strength of the prominences is derived by assuming the oscillations to be kink-mode oscillations. The value for the prominence $P_{1U}$ is roughly consistent with the result derived by an independent method, i.e. a traveling wave.

It is noted that all three layers of the first prominence were oscillating collectively. However, the prominence seismology in this paper is based on a single-segment configuration for the filament, which may introduce uncertainties in the estimate of the magnetic field. Considering that there exist thread-thread interactions \citep{Zhou2017}, it would be interesting to investigate their influence on the estimate of the prominence magnetic field in the future.

\begin{table}
    \centering
     \caption{The measured and derived parameters of the prominences.}
     \medskip
    \begin{tabular}{|c|c|c|c|}
    \hline
      Prominence & Length & Period &  Derived  \\
       & (Mm) &  (min) &  magnetic field (G)  \\
      
      \hline
       P$_{1L}$ & 220  & 17 $\pm$ 1.5 & 14.4 $\pm$ 1.3 \\
       P$_{1M}$ & 260 &  14 $\pm$ 1.2 &  20.1 $\pm$ 1.8 \\
       P$_{1U}$ & 340 & 22 $\pm$ 2.0 & 16.3 $\pm$ 1.6 \\
       P$_{2}$ & 210 &  16 $\pm$ 1.3 & 14.4 $\pm$ 1.2 \\
       \hline
    \end{tabular}
   
    \label{table}
\end{table}

\vspace{1cm}

\noindent {\it Acknowledgments:}
We thank both reviewers for their comments and suggestions. The authors thank the open data policy of the SDO team. PD thanks the CSIR, New Delhi for providing the Research Fellowship. PFC was supported by the National Key Research and
Development Program of China (2020YFC2201200) and NSFC (11961131002), RC acknowledges the support from Bulgarian Science Fund under Indo-Bulgarian bilateral project, DST/INT/BLR/P-11/2019. YJM was supported by the Basic Science Research Program through the NRF funded by the Ministry of Education (NRF-2019R1A2C1002634).  We thank Dr. Manuel Luna for discussions on timeslice analysis.

\bibliographystyle{model5-names}
\biboptions{authoryear}
\bibliography{reference}

\end{document}